\pdfoutput=1

\documentclass[10pt,letterpaper]{article}

\usepackage{subfig}
\usepackage{colortbl}
\usepackage{balance}

\usepackage{mathptmx}

\usepackage{url}
\usepackage[multiple,flushmargin]{footmisc}

\usepackage{booktabs,ctable,footnote}
\usepackage{multirow}

\usepackage{newalg}
\usepackage{amsfonts}

\RequirePackage{ifpdf}
\ifpdf%
\RequirePackage{pdflscape}
\DeclareGraphicsExtensions{.pdf,.png,.jpg}
\else
\RequirePackage{lscape}
\DeclareGraphicsExtensions{.eps,.ps}
\RequirePackage[ps2pdf]{thumbpdf}
\fi

\newcommand{\acro}[1]{\textsc{\lowercase{#1}}}

\newcommand{\fn}[1]{\footnote{\scriptsize #1}}
\newcommand{\ar}[1]{{\small \textsf{\uppercase{#1}}}}
\newcommand{\s}[1]{\textbf{\texttt{\small #1}}}

\newcommand{\ms}[1]{\mathbb{#1}}
\newcommand{\B}{\ms{B}}

\newcommand{\otoprule}{\midrule[\heavyrulewidth]}
\newcolumntype{+}{>{\global\let\currentrowstyle\relax}}
\newcolumntype{-}{>{\currentrowstyle}}
\newcommand{\rowstyle}[1]{\gdef\currentrowstyle{#1} #1\ignorespaces}

\begin{document}

\title{A Grey-Box Approach to Automated Mechanism Design}

\author{Jinzhong Niu$^{1,2}$, Kai Cai$^{1}$, and Simon Parsons$^{1,3}$
\\
\\
$^{1}$Department of Computer Science\\
The Graduate School and University Center, The City University of New York\\
365, Fifth Avenue, New York, NY 10016, USA\\
\texttt{\{jniu,kcai\}@gc.cuny.edu}
\and
$^{2}$School of Computer Science, University of Birmingham\\
Edgbaston, Birmingham B15 2TT, UK
\and
$^{3}$Department of Computer and Information Science\\
Brooklyn College, The City University of New York\\
2900 Bedford Avenue, Brooklyn, NY 11210\\
\texttt{parsons@sci.brooklyn.cuny.edu}
}

\maketitle


\begin{abstract}

Auctions play an important role in electronic commerce, and have been used to solve problems in distributed computing. Automated approaches to designing effective auction mechanisms are helpful in reducing the burden of traditional game theoretic, analytic approaches and in searching through the large space of possible auction mechanisms. This paper presents an approach to automated mechanism design (\acro{AMD}) in the domain of double auctions. We describe a novel parameterized space of double auctions, and then introduce an evolutionary search method that searches this space of parameters. The approach evaluates auction mechanisms using the framework of the TAC Market Design Game and relates the performance of the markets in that game to their constituent parts using reinforcement learning. Experiments show that the strongest mechanisms we found using this approach not only win the Market Design Game against known, strong opponents, but also exhibit desirable economic properties when they run in isolation.

\end{abstract}

\section{Introduction}
\label{sec:intro}

In the Internet era, ecommerce has grown and flourished. The greater amount of available information, the lower cost of communication, and other reductions in economic friction makes the world `flatter' than ever before, promoting automated marketplaces and the adoption of autonomous agents in ecommerce \cite{greenwal-jennings-stone-03-iis-agents.and.markets,kephart-02-pnas-software.agents}. In financial markets, traders have continuously turned to automated algorithmic trading services to deal with faster transactions and more complex market dynamics \cite{schartz:francioni:weber:06:equity-trader-course}. According to an article from \textit{The Economist} \cite{070623-economist-algo.trading}, algorithmic trading accounted for a third of all share trades in the United States, and Aite Group, a consultancy, reported that the figure will reach 50\% by 2010. News and events usually affect market predictions and lead to high price volatility, which in turn creates opportunities for arbitrage between markets. Unpredictable dynamics and complex linkages between markets make more robust, efficient market mechanisms very desirable.

Online auction sites like eBay provide a way for consumers to buy a wide range of items. Since its establishment in 1995, eBay has expanded into dozens of countries and now makes billions of dollars each year. The auction mechanisms used by eBay and other successful auction sites however are not perfect. For example, an eBay auction typically finishes at a fixed time, allowing a bidder to bid only moments before the auction terminates and steal a deal from bidders who would offer higher prices if given the chance \cite{greenwald06e-markets}. This means a loss of revenue for both sellers and eBay. Another issue, and one to which many researchers have paid much attention, is that eBay runs many \emph{simultaneous sequential auctions}  \cite{gerding-dyj-07-aamas-bidding.in.concurrent.auctions,juda-parkes-06-ec-sequential.ebay.auctions}. In other words, on eBay, hundreds, even thousands, of on-going auctions may sell the same kind of goods. It is difficult for a potential buyer to select an auction that would result in the lowest winning bid. As a result, a successful bid in one auction may be lower than a failed bid in another, leading to complaints from both sellers and bidders, lower efficiency of the auctions and, in time, less revenue for eBay.

Electronic auctions have also been used to sell things that are not goods in a traditional sense. For example, search engines like Google and Yahoo, in the role of  publisher, typically use auctions to select and show relevant advertisements along with search results on their web sites. For each keyword-based search query, an \emph{ad auction} is run to select bids from advertisers. Each selected advertiser provides an ad to display in one of a certain number of \emph{ad positions} on the search result page. Better positions, which draw more attention from users, are allotted to advertisers that bid higher. An advertiser usually pays on a \emph{per-click} basis rather than on the \emph{per-impression} basis in the traditional media. Publishers have commonly used variants of an auction mechanism called the \emph{generalized second-priced auction} to determine winning bids from advertisers and their ad positions. Although ad auctions generate many dollars in income each year for these companies, how to analyze the current practices and design more effective ad auction mechanisms is still a major concern. For instance, Lahaie and Pennock \cite{lahaie:pennock:07:revenue-analysis} compared the ranking rule used by Yahoo --- based on the prices of bids --- and that used by Google --- based on the expected profits of bids to Google, and concluded that neither rule consistently outperforms the other.

All these scenarios from e-commerce challenge the designers of electronic auction mechanism to design more desirable mechanisms. This opens up new lines of research in computer science, such as inventing new algorithms for deciding the winning bid in auctions \cite{lehmann-muller-sandholm-06-bookchap-wdp}, deciding how best to bid in multiple auctions \cite{schwartz:byrne:colaninno:08:competition}, and how to build the software infrastructure to run such auctions \cite{niu-cpgmmps-08-aamas-jcat}.

The Internet also significantly boosts the adoption of distributed computing, in particular \emph{agent-based computing}. A major tool for multi-agent system designers has been \emph{game theory} (\acro{gt}). \acro{gt} provides a framework for studying strategic, interacting individuals and solution concepts --- usually various equilibria --- with the assumption of the rationality of individuals. \acro{gt} thus helps to compare the outcomes of an interaction mechanism to the optimal ones in theory, but it does not give a dynamic model that explains how to reach optimal outcomes, nor presents much guidance on how to maximize global outcomes when some agents in the system are not as rational as presumed.

\emph{Auction mechanisms} are an ideal candidate to provide this missing model. The effectiveness of auction mechanisms in the real world and the similarity between an auction and a multi-agent system --- both involving multiple self-interested individuals and concerning certain global outcomes --- have led to various market-based approaches to multi-agent coordination and resource allocation problems in cluster and grid computing environments \cite{horling-lesser-05-ker-organizational.paradigms,stober-neumann-08-ecr-greedex,yeoB06taxonomy}. These approaches have demonstrated superior performance than those pre-existing, non-market solutions, in terms of a combination of performance, scalability, and reliability. However, the market mechanisms adopted by these approaches are usually selected arbitrarily or based on certain heuristics. It is unknown whether these market mechanisms are optimal solutions, or whether there are better options.

\section{A new approach}
\label{sec:strategic-decision-making}

\subsection{Automated mechanism design}
\label{sec:strategic-decision-making:amd}

Facing the challenges in both electronic commerce and market-based control, we need to solve the following problem: \emph{Given a certain set of restrictions and desired outcomes, how can we design a good, if not optimal, auction mechanism; or when the restrictions and goals alter, how can the current mechanism be improved to handle the new scenario?}

The traditional answer to this question has been in the domain of auction theory \cite{krishna-02-book-auction.theory}. A mechanism is designed by hand, analyzed theoretically, and then revised as necessary. The problems with the approach are exactly those that dog any manual process --- it is slow, error-prone, and restricted to just a handful of individuals with the necessary skills and knowledge. In addition, there are classes of commonly used mechanism, such as the double auctions that we discuss here, which are too complex to be analyzed theoretically, at least for interesting cases \cite{walsh-02-gtdt-analyzing.complex.strategic.interactions}.

Automated mechanism design (\acro{amd}) aims to overcome the problems of the manual process by designing auction mechanisms automatically. \acro{amd} considers design to be a search through some space of possible mechanisms. For example, Cliff \cite{cliff-01-tr-evolution.of.market.mech} and Phelps \textit{et al.} \cite{phelps02coev.mechanism.design,phelps03optimizing.pricing.rules} explored the use of evolutionary algorithms to optimize different aspects of the continuous double auction. Around the same time, Conitzer and Sandholm \cite{conitzer-sandholm-03-icec-amd.complexity} were examining the complexity of building a mechanism that fitted a particular specification.

These different approaches were all problematic. The algorithms that Conitzer and Sandholm considered dealt with exhaustive search, and naturally the complexity was exponential. In contrast, the approaches that Cliff and Phelps \textit{et al.} pursued were computationally more appealing, but gave no guarantee of success and were only searching tiny sections of the search space for the mechanisms they considered. As a result, one might consider the work of Cliff and Phelps \textit{et al.}, and indeed the work we describe here, to be what Conitzer and Sandholm \cite{conitzer-sandholm-07-ijcai-incremental} call ``incremental'' mechanism design, where one starts with an existing mechanism and incrementally alters parts of it, aiming to iterate towards an optimal mechanism. Similar work, though work that uses a different approach to searching the space of possible mechanisms has been carried out by \cite{vorobeychik-reeves-wellman-07-uai-constrained.amd} and has been applied to several different mechanism design problems \cite{schvartzman-wellman-09-amec-exploring.large.strategy.spaces}.

The problem with taking the automated approach to mechanism design further is how to make it scale --- though framing it as an incremental process is a good way to look at it, it does not provide much practical guidance about how to proceed. Our aim in this paper is to provide more in the way of practical guidance, showing how it is possible to build on a previous analysis of the most relevant components of a complex mechanism in order to set up an automated mechanism design problem, and then describing one approach to solving this problem.

\subsection{CAT games}
\label{sec:strategic-decision-making:cat}

We set our work within the context of the Trading Agent Competition Market Design game, also known as the \acro{cat} game. This competition, which ran for the last three years, asks entrants to design a market for a set of automated traders which are based on standard algorithms for buying and selling in a double auction, including \acro{zi-c} \cite{gode-sunder-93-jpe-zi}, \acro{zip} \cite{cliff-97-tr-zip}, \acro{re} \cite{erev98predicting}, and \acro{gd} \cite{gjerstad-dickhaut-98-geb-gd}. The game is broken up into a sequence of \emph{days}, and each day every trader picks a market to trade in, using a market selection strategy that models the situation as an $n$-armed bandit problem \cite[Section 2]{sutton-barto-98-book-rl}. Markets are allowed to charge traders in a variety of ways and are scored based on the number of traders they attract (market share), the profits that they make from traders (profit share), and the number of successful transactions they broker relative to the total number of shouts placed in them (transaction success rate). Full details of the game can be found in \cite{cai-gmnpp-09-tr-cat.overview}.

We picked the \acro{cat} game as the basis of our work for four main reasons. First, the double auctions that are the focus of the design are a widely used mechanism. Second, the competition is run using an open-source software package called \acro{jcat} which is a good basis for implementing our ideas. Third, after three years of competition, a number of specialists have been made available by their authors, giving us a library of mechanisms to test against. Fourth, there have been a number of publications that analyze different aspects of previous entrants, giving us a good basis from which to start searching for new mechanisms.

\subsection{Towards a grey-box approach}
\label{sec:strategic-decision-making:white+black}

Particularly helpful is the prior work of Niu \textit{et al.} \cite{niu-cpgm-08-aamas-cats,niu-cmp-08-iat-eco}. \cite{niu-cpgm-08-aamas-cats} is an analysis of the 2007 \acro{cat} competition, which identifies a number of different components of the double auction market, along with different implementations of these components proposed by the game entrants. \cite{niu-cmp-08-iat-eco} complements this analysis with a description of a large number of simulations of competitions between the specialists from the 2007 \acro{cat} game, systematically identifying how different specialists perform both in multi-market games, and in games between pairs of specialists. Together these two papers mirror the black-box and white-box analyses from software engineering. \cite{niu-cpgm-08-aamas-cats} provides a white-box analysis, looking inside each market in order to identify which components it contains, and relating the performance of each market to the operation of its components. \cite{niu-cmp-08-iat-eco} provides a black-box analysis, which ignores the detail of the internal components of each market, but providing a much more extensive analysis of how the specialists perform.

These analyses make a good combination for examining the strengths and weaknesses of specialists. The white-box approach is capable of relating the internal design of a strategy to its performance and revealing which part of the design may cause vulnerabilities, but it requires internal structure and involves manual examination. The black-box approach does not rely upon the accessibility of the internal design of a strategy. It can be applied to virtually any strategic game, and is capable of evaluating a design in many more situations. However, the black-box approach tells us little about what may have caused a strategy to perform poorly and provides little in the way of hints as to how to improve the strategy. It is desirable to combine these two approaches in order to benefit from the advantages of both. Following the \acro{ga}-based approach to trading strategy acquisition and auction mechanism design in \cite{cliff-01-tr-evolution.of.market.mech,phelps06automatic-strategy-acquisition,phelps03optimizing.pricing.rules}, we propose what we call a \emph{grey-box} approach to automated mechanism design that solves the problem of automatically creating a complex mechanism by searching a structured space of auction components.

In other words, we concentrate on the components of the mechanisms (as in the white-box approach), but take a black-box view of the components, evaluating their effectivenesses by looking at their performance against that of their peers.

More specifically, we view a market mechanism as a combination of auction rules, each as an atomic building block. We consider the problem: \emph{how can we find a combination of rules that is better than any known combination according to a certain criterion, based on a pool of existing building blocks?} The black-box analysis in \cite{niu-cmp-08-iat-eco} maintains a population of strategies and evolves them generation by generation based on their fitnesses. Here we intend to follow a similar approach, maintaining a population of components or building blocks for strategies, associating each block with a \emph{quality score}, which reflects the fitnesses of auction mechanisms using this block, exploring the part of the space of auction mechanisms that involves building blocks of higher quality, and keeping the best mechanisms we find.

\section{Grey-box AMD}
\label{sec:greybox}

Having sketched our approach at a high level, we now look in detail at how it can be applied in the context of the \acro{cat} game.

\subsection{A search space of double auctions}
\label{sec:greybox:gda}

The first issues we need to address are \emph{what composite structure is used to represent auction mechanisms?} and \emph{where can we obtain a pool of building blocks?}

Viewing an auction as a structured mechanism is not a new idea. Wurman \textit{et al.} \cite{wurman01parametrization} introduced a conceptual, parameterized view of auction mechanisms. Niu \textit{et al.} \cite{niu-cpgm-08-aamas-cats} extended this framework for auction mechanisms competing in \acro{cat} games and provided a classification of entries in the first \acro{cat} competition that was based on it. The extended framework includes multiple intertwined components, or \emph{policies}, each regulating one aspect of a market. We adopt this framework, include more candidates for each type of policy and take into consideration parameters that are used by these policies.

These policies are either inferred from the literature \cite{mccabe93upda}, or from our previous work \cite{niu-cmp-08-iat-eco,niu-cpgm-08-aamas-cats,niu-cps-06-aamas-pricing.fluctuation}, or contributed by entrants to the \acro{cat} competitions. These policies, each as a building block, form a solid foundation for the grey-box approach.

\begin{sidewaysfigure}
\begin{center}
\resizebox{.85\linewidth}{!}{
  \includegraphics{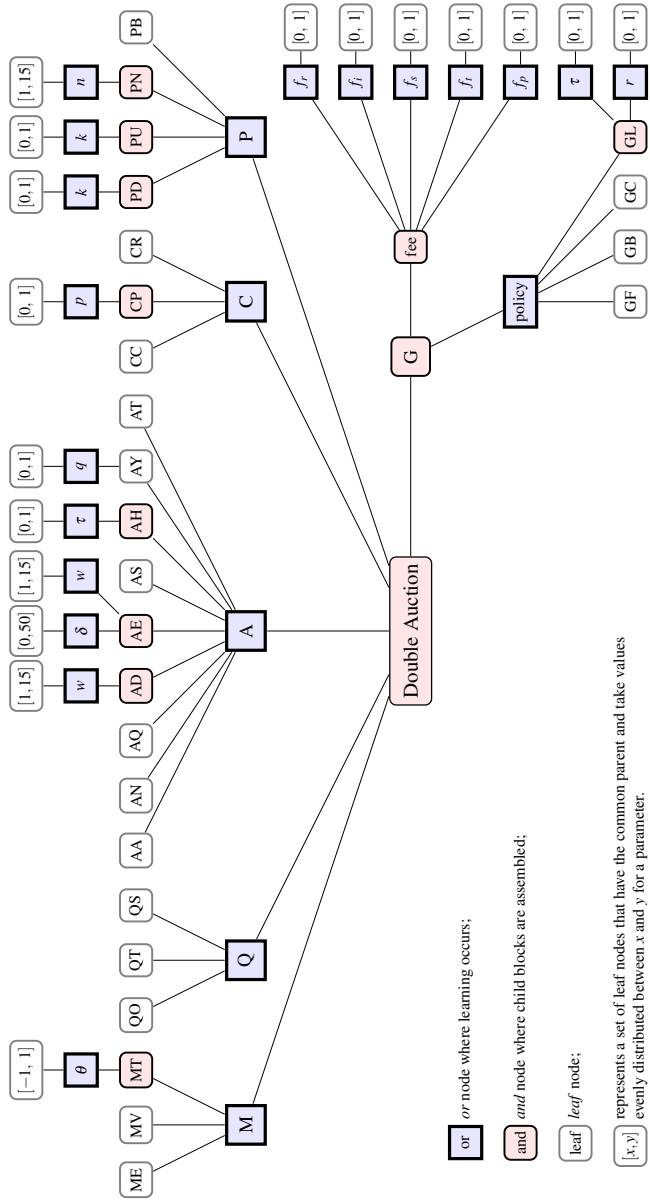}
}
\caption{The search space of double auctions modeled as a tree, discussed in details in Section~\ref{sec:greybox}.}
\label{fig:auction-space-tree}
\end{center}
\end{sidewaysfigure}

Figure~\ref{fig:auction-space-tree} illustrates the building blocks as a tree structure which we describe after we review the blocks themselves.
%
%
We describe the different types of policies in details below and discuss how we search the space based on the tree structure in the next section.

\subsubsection{Matching policy}
\label{sec:greybox:gda:matching}

Matching policies, denoted as \ar{M} in Figure~\ref{fig:auction-space-tree}, define how a market matches shouts made by traders.
\emph{Equilibrium matching} (\ar{me}) is the most commonly used matching policy \cite{mccabe93upda,wurman-98-dss-flexible.das}. The offers made by traders form the \emph{reported demand and supply}, which is usually different from the \emph{underlying demand and supply}, and are determined by traders' private values and unknown to the market, since traders are assumed to be profit-seeking and make offers deviating from their private values. \ar{me} clears the market at the \emph{reported} equilibrium price and matches intra-marginal asks (offers to sell) with intra-marginal bids (offers to buy) --- with an intersecting demand and supply, the shouts on the left of the intersection (the equilibrium point) and their traders are called \emph{intra-marginal} since they can be matched and make profit, while those on the right are called \emph{extra-marginal}. It is worth mentioning that a shout, or a trader, that appears to be intra-marginal or extra-marginal in the reported demand and supply may not be so in the underlying demand and supply.
\emph{Max-volume matching} (\ar{mv}) aims to increase transaction volume based on the observation that a high intra-marginal bid can match with a lower extra-marginal ask, though with a profit loss for the buyer. It does so to realize the maximal transaction volume that is possible.
A generic, parameterized, matching policy can be defined to include \ar{me} and \ar{mv} as two special cases. This policy, denoted as \ar{mt}, uses a parameter, $\theta$, which can be any value in $[-1,1]$. When $\theta$ is $-1$, \ar{mt} does not match any shout; when $\theta$ is $0$, \ar{mt} becomes \ar{me}; and when $\theta$ is $1$, \ar{mt} becomes \ar{mv}. For any other values of $\theta$, \ar{mt} tries to realize a transaction volume that is proportional to $0$ and those realized in \ar{me} and \ar{mv}.

\subsubsection{Quote policy}
\label{sec:greybox:gda:quoting}

Quote policies, denoted as \ar{Q} in Figure~\ref{fig:auction-space-tree}, determine the quotes issued by markets. Typical quotes are ask and bid quotes, which respectively specify the upper bound for asks and the lower bound for bids that may be placed in a \emph{quote-driven} market. \emph{Two-sided quoting}\fn{The name follows \cite{mccabe93upda} since either  quote depends on information on both the ask side and the bid side.} (\ar{qt}) defines the ask quote as the minimum of the lowest tentatively matchable bid and lowest unmatchable ask, and defines the bid quote as the the maximum of the highest tentatively matchable ask and highest unmatchable bid.
%
%
\emph{One-sided quoting} (\ar{qo}) is similar to \ar{qt}, but considers only the standing shouts closest to the reported equilibrium price from the unmatchable side. When the market is cleared continuously (see below), \ar{qo} is identical to \ar{qt}, but otherwise forms a possibly looser restriction on placing shouts.
\emph{Spread-based quoting} (\ar{qs}) extends \ar{qt} to maintain a higher ask quote and a lower bid quote for use with \ar{mv}. With \ar{qs}, when the ask quote is lower than the bid quote, the former is set somewhere above their average and the latter below the average, and the spread between the two is a fixed value. \ar{qs} helps relax the constraint put on shouts with too low an ask quote and too high a bid quote.

\subsubsection{Shout accepting policy}
\label{sec:greybox:gda:accepting}

Shout accepting policies, denoted as \ar{A} in Figure~\ref{fig:auction-space-tree}, judge whether a shout made by a trader should be permitted in the market.
\emph{Always accepting} (\ar{aa}) accepts any shout, and \emph{never accepting} (\ar{an}) does the opposite.
\emph{Quote-beating accepting} (\ar{aq}) allows only those shouts that are more competitive than the corresponding market quote. This has been commonly used in both experimental settings and real stock markets, and is sometimes called the ``New York Stock Exchange rule'' since that market adopts it.
%
%
\emph{Self-beating accepting} (\ar{as}) accepts all first-time shouts but only allows a trader to modify its standing shout with a more competitive price.
%
\emph{Equilibrium-beating accepting} (\ar{ae}) learns an estimate of the equilibrium price based on the past transaction prices in a sliding window, and requires bids to be higher than the estimate and asks to be lower. \ar{ae} uses a parameter, $w$, to specify the size of the sliding window in terms of the number of transactions, and a second parameter, $\delta$, which can be added to the estimate to relax the restriction on shouts. This policy was suggested in \cite{niu-cps-06-aamas-pricing.fluctuation} and found to be effective in reducing transaction price fluctuation and increasing allocative efficiency in markets populated by \acro{zi-c} traders.
A variant of \ar{ae}, denoted as \ar{ad} and introduced by the \s{PSUCAT} team in the first \acro{cat} competition, uses the standard deviation of transaction prices in the sliding window rather than a constant $\delta$ to relax the restriction on shouts.
\emph{History-based accepting} (\ar{ah}) is derived from the \acro{gd} trading strategy \cite{gjerstad-dickhaut-98-geb-gd} and reported to be a crucial component of one particular strong market mechanism for \acro{cat} games \cite{niu-cmp-08-iat-eco}. \acro{gd} computes how likely a given offer is to be matched, based on the history of previous shouts, and \ar{ah} uses this to accept only shouts that will be matched with probability no lower than a specified threshold, $\tau\in [0,1]$.
\emph{Transaction-based accepting} (\ar{at}) tracks the most recently matched asks and bids, and uses the lowest matched bid and the highest matched ask to restrict the shouts to be accepted. In a clearing house (\acro{ch}) \cite{friedman-rust-93-book-dam}, the two bounds are expected to be close to the estimate of equilibrium price in \ar{ae}, while in a continuous double auction (\acro{cda}), \ar{at} may produce much looser restriction since extra-marginal shouts may steal a deal.
\emph{Shout type-based accepting} (\ar{ay}) allows shouts based merely on their types, i.e. asks or bids. This mimics the continuum of auctions presented in \cite{cliff-01-tr-evolution.of.market.mech}, including retailer markets where only sellers shout, procurement auctions where only buyers shout, as well as general double auctions.

\subsubsection{Clearing condition}
\label{sec:greybox:gda:clearing}

Clearing conditions, denoted as \ar{C} in Figure~\ref{fig:auction-space-tree}, define when to clear the market and execute transactions between matched asks and bids.
\emph{Continuous clearing} (\ar{cc}) attempts to clear the market whenever a new shout is placed.
\emph{Round clearing} (\ar{cr}) clears the market after all traders have submitted their shouts. This was the original clearing policy in \acro{nyse}, but was replaced by \ar{cc} later for faster transactions and higher volumes. With \ar{cc}, an extra-marginal trader may have more chance to steal a deal and get matched.
\emph{Probabilistic clearing} (\ar{cp}) clears the market with a predefined probability, $p$, whenever a shout is placed. It thus defines a continuum of clearing rules with \ar{cr} ($p=0$) and \ar{cc} ($p=1$) being the two ends.

\subsubsection{Pricing policy}
\label{sec:greybox:gda:pricing}

Pricing policies, denoted as \ar{P} in Figure~\ref{fig:auction-space-tree}, set transaction prices for matched ask-bid pairs. The decision making may involve only the prices of the matched ask and bid, or more information including market quotes.
\emph{Discriminatory $k$-pricing} (\ar{pd}) sets the transaction price of a matched ask-bid pair at some point in the interval between their prices. The parameter $k\in [0,1]$ controls which point is used and usually takes value $0.5$ to avoid a bias in favor of buyers or sellers.
\emph{Uniform $k$-pricing} (\ar{pu}) is similar to \ar{pd}, but sets the transaction prices for all matched ask-bid pairs at same point between the ask quote and the bid quote. A transaction price set by \ar{pu} may or may not fall into the range between the matched ask and bid, depending upon the matching policy and the quote policy in the auction mechanism, When it falls outside, whichever of the ask and the bid is closer to the computed transaction price will be used as the final transaction price.
\emph{$n$-pricing} (\ar{pn}) was introduced in \cite{niu-cps-06-aamas-pricing.fluctuation}, and sets the transaction price as the average of the latest $n$ pairs of matched asks and bids. If the average falls out of the price interval between the ask and bid to be matched, the nearest end of the interval is used. This policy can help reduce transaction price fluctuation and has little impact on allocative efficiency.
\emph{Side-biased pricing} (\ar{pb}) is basically \ar{pd} with an internal $k$ dynamically adjusted so as to split the profit in favor of the side on which fewer shouts exist. Thus the more that asks outnumber bids in the current market, the closer $k$ is set to 0.

\subsubsection{Charging policy}
\label{sec:greybox:gda:charging}

Charging policies, denoted as \ar{G} in Figure~\ref{fig:auction-space-tree}, determine the charges imposed by a market. This is typically not an issue in research on auctions in isolation, but would affect the selection of markets by  traders directly in an environment of multiple competing markets, as in \acro{cat} games.
\emph{Fixed charging} (\ar{gf}) sets charges at a specified fixed level.
\emph{Bait-and-switch charging} (\ar{gb}) makes a market cut its charges until it captures a certain market share, and then slowly increases charges to increase profit. It will adjust its charges downward again if its market share drops below a certain level.
\emph{Charge-cutting charging} (\ar{gc}) sets the charges by scaling down the lowest charges imposed on the previous day, based on the observation that traders prefer markets with lower charges.
\emph{Learn-or-lure-fast charging} (\ar{gl}) adapts charges towards some target following the scheme used by the \acro{zip} trading strategy \cite{cliff-97-tr-zip}. If the market using this policy believes that the traders are still exploring among markets and have yet to find a good one to trade, the market would adapt charges towards 0 to lure traders to join and stay; otherwise it learns from the charges of the most profitable market. \ar{gl} uses an exploring monitor component to determine whether traders are exploring or not. A simple exploring monitor, for example, examines the daily distribution of market shares of specialists. If the distribution is flat, the traders are considered exploring, and not otherwise. This is based on the observation that traders all tend to go to the best market and cause an imbalanced distribution. To implement this, the degree of flatness of the distribution --- the standard deviation of the distribution relative to the mean of the distribution --- is compared to a threshold $\tau \in [0,1]$. And similar to \acro{zip}, \ar{gl} uses a learning rate parameter, $r$, to control how fast the market adapts its charges.

All these charging policies require an initial set of fees on different activities, including fee on registration, fee on information, fee on shout, fee on transaction, and fee on profit, denoted as $f_r$, $f_i$, $f_s$, $f_t$, and $f_p$ respectively in Figure~\ref{fig:auction-space-tree}.

\subsubsection{A tree model}
\label{sec:greybox:gda:tree}

The tree model of double auctions in Figure~\ref{fig:auction-space-tree} illustrates how building blocks are selected and assembled level by level. There are \emph{and} nodes, \emph{or} nodes, and \emph{leaf} nodes in the tree. An \emph{and} node, rounded and filled, combines a set of building blocks, each represented by one of its child nodes, to form a compound building block. The root node, for example, is an \emph{and} node to assemble policies, one on each aspect described above, to obtain a complete auction mechanism. An \emph{or} node, rectangular and filled, represents the decision making of selecting a building block from the candidates represented by the child nodes of the \emph{or} node based on their quality scores. This selection occurs not only for those major aspects of an auction mechanism, i.e. \ar{m}, \ar{q}, \ar{a}, \ar{p}, \ar{c}, and \ar{g} (at \ar{g}'s child node of `policy' in fact), but also for minor components, for example, a learning component for an adaptive policy (as what Phelps \textit{et al.} does regarding a trading strategy \cite{phelps06automatic-strategy-acquisition}), and for determining optimal values of parameters in a policy, like $\theta$ in \ar{mt} and $k$ in \ar{pd}. A \emph{leaf} node represents an atomic block that can either be for selection at its \emph{or} parent node or be further assembled into a bigger block by its \emph{and} parent node. A special type of \emph{leaf} node in Figure~\ref{fig:auction-space-tree} is that with a label in the format of $[x,y]$. Such a \emph{leaf} node is a convenient representation of a set of \emph{leaf} nodes  that have a common parent --- the parent of this special \emph{leaf} node --- and take values evenly distributed between $x$ and $y$ for the parameter labeled at the parent node.

\emph{or} nodes contribute to the variety of auction mechanisms in the search space and are where exploitation and exploration occur. We model each \emph{or} node as an $n$-armed bandit learner that chooses among candidate blocks, and use the simple softmax method \cite[Section 2.3]{sutton-barto-98-book-rl} to solve this learning problem. The same solution is adopted in designing the market selection strategy for trading agents in \acro{cat} games \cite{niu-cps-07-tada-market.competition}.\footnote{However the two scenarios may need different parameter values. The market selection scenario should favor choices that give a good profit --- a cumulative measure --- while here we require effective exploration to find a good mechanism in the foreseeable future --- a one-time concern.}

\subsection{The \textsc{Grey-Box-AMD} algorithm}
\label{sec:greybox:algorithm}

\newcommand{\EM}{\ms{EM}}
\newcommand{\HOF}{\ms{HOF}}
\newcommand{\FM}{\ms{FM}}
\newcommand{\SM}{\ms{SM}}
\newcommand{\M}{\ms{M}}
\renewcommand{\S}{\ms{S}}
\newcommand{\G}{\ms{G}}
\newcommand{\X}{\ms{X}}
\newcommand{\Y}{\ms{Y}}
\newcommand{\const}[1]{\texttt{\textsc{#1}}}
\newcommand{\kw}[1]{\textbf{#1}}

Given a set of building blocks, $\B$, and a set of fixed markets, $\FM$, as targets to beat, we define the skeleton of the grey-box algorithm below:

\medskip

\begin{algorithm}{Grey-Box-AMD}{\B, \FM}
\HOF \= \{\} \\
\begin{FOR}{s \= 1 \TO \const{num\_of\_steps}}
  G \= \CALL{Create-Game}() \\
  \SM \= \{\} \\
  \begin{FOR}{m \= 1 \TO \const{num\_of\_samples}}
    M \= \CALL{Create-Market}() \\
    \begin{FOR}{t \= 1 \TO \const{num\_of\_policytypes}}
    B \= \CALL{Select}(\B_t, 1) \\
    \CALL{Add-Block}(M, B)
    \end{FOR} \\
    \SM \= \SM \cup \{M\}
  \end{FOR} \\
  \EM \= \CALL{Select}(\HOF, \const{num\_of\_hof\_samples})) \\
  \CALL{Run-Game}(G, \FM\cup\EM\cup\SM) \\
  \begin{FOR}{\EACH M \IN \EM \cup \SM}
    \CALL{Update-Market-Score}(M, \CALL{Score}(G, M)) \\
    \begin{IF}{M\ \kw{not} \IN \HOF}
        \HOF \= \HOF \cup \{M\}
    \end{IF} \\
    \begin{IF}{\const{capacity\_of\_hof} < |\HOF|}
    \HOF \= \HOF  - \{\CALL{Worst-Market}(\HOF)\}
    \end{IF} \\
    \begin{FOR}{\EACH B \text{used by} M}
      \CALL{Update-Block-Score}(B, \CALL{Score}(G, M))
    \end{FOR}
  \end{FOR}
\end{FOR} \\
\RETURN \HOF
\end{algorithm}

The \textsc{Grey-Box-AMD} algorithm runs a certain number of steps (\const{num\_of\_steps} in Line 2). At each step, a single \acro{cat} game is created (\textsc{Create-Game}() in Line 3) and a set of markets are prepared for the game. This set of markets includes all markets in $\FM$, a certain number (\const{num\_of\_samples} in Line 5) of markets sampled from the search space, denoted as $\SM$, and a certain number (\const{num\_\-of\_\-hof\_samples} in Line 11) of markets, denoted as $\EM$, chosen from a Hall of Fame, $\HOF$. All these markets are put into the game, which is run to evaluate the performance of these markets (\textsc{Run-Game}($G$, $\FM\cup\EM\cup\SM$) in Line 12). $\HOF$ has a fixed capacity, \const{capacity\_of\_hof}, and maintains markets that performed well in games at previous steps in terms of their average scores across games they participated. $\HOF$ is empty initially, updated after each game, and returned in the end as the result of the grey-box process.

Each market in $\SM$ is constructed based on the tree model in Figure~\ref{fig:auction-space-tree}. After an `empty' market mechanism, $M$, is created (\textsc{Create-Market}() in Line 6), building blocks can be incorporated into it (\textsc{Add-Block}($M$,$B$) in Line 9, where $B\in\B$). \const{num\_of\_policytypes} in Line 7 defines the number of different policy types, and from each group of policies of same type, denoted as $\B_t$ where $t$ specifies the type, a building block is chosen for $M$ (\textsc{Select}($\B_t$, $1$) in Line 8). For simplicity, this algorithm illustrates only what happens to the \emph{or} nodes at the high level, including \ar{M}, \ar{Q}, \ar{A}, \ar{C}, and \ar{P}. Markets in $\EM$ are chosen from $\HOF$ in a similar way (\textsc{Select}($\HOF$, \const{num\_of\_hof\_samples}) in Line 11).

After a \acro{cat} game, $G$, completes at each step, the game score of each participating market $M\in\SM\cup\EM$, \textsc{Score}($G$, $M$), is recorded and the game-independent score of $M$, \textsc{Score}($M$),  is updated (\textsc{Update-Market-Score}($M$, \textsc{Score}($G$, $M$)) in Line 14). If $M$ is not currently in $\HOF$ and \textsc{Score}($M$) is higher than the lowest score of markets in $\HOF$, it replaces that corresponding market (\textsc{Worst-Market}($\HOF$) in Line 18).

\textsc{Score}($G$, $M$) is also used to update the quality score of each building block used by $M$ (\textsc{Update-Block-Score}($B$, \textsc{Score}($G$, $M$)) in Line 20). Both \textsc{Update-Market-Score} and \textsc{Update-Block-Score} calculate respectively game-independent scores of markets and quality scores of building blocks by averaging feedback \textsc{Score}($G$, $M$) over time. Because choosing building blocks occurs only at \emph{or} nodes in the tree, only child nodes of an \emph{or} node have quality scores and receive feedback after a \acro{cat} game. Initially, quality scores of building blocks are all $0$, so that the probabilities to choose them are even. As the exploration proceeds, fitter blocks score higher and are chosen more often to construct better mechanisms.

\section{Experiments}
\label{sec:experiments}

This section describes the experiments that are carried out to acquire auction mechanisms using the grey-box approach.

\subsection{Experimental setup}
\label{sec:experiments:setup}

We extended \acro{jcat} with the parameterized framework of double auctions and all the individual policies described in Section~\ref{sec:greybox:gda}. To reduce the computational cost, we eliminated the exploration of charging policies by focusing on mechanisms that impose a fixed charge of $10\%$ on trader profit, which we denote as $\ar{gf}_{0.1}$. Analysis of \acro{cat} games \cite{niu-cmp-08-iat-eco} and what entries have typically charged in actual \acro{cat} competitions, especially in the latest two events, suggest that such a charging policy can be a reasonable choice to avoid losing either intra-marginal or extra-marginal traders. Even with this cut-off, the search space still contains more than $1,200,000$ different kinds of auction mechanisms, due to the variety of policies on aspects other than charging and the choices of values for parameters.

The experiments that we ran to search the space each last 200 steps. At each step, we sample two auction mechanisms from the space, and run a \acro{cat} game to evaluate them against four fixed, well known, mechanisms plus two mechanisms that performed well at previous steps and are members of the Hall of Fame. The scores of the sampled and Hall of Fame mechanisms are used as feedback for every building block that an individual mechanism uses and is associated with a quality score.

To sample auction mechanisms, the softmax exploration method used by \emph{or} nodes starts with a relatively high temperature so as to explore randomly, then gradually cools down, and eventually maintains a temperature that guarantees a non-negligible probability of choosing even the worst action any time. After all, our goal in the grey-box approach is not to converge quickly to a small set of mechanisms, but to explore the space as broadly as possible and avoid being trapped in local optima.

The fixed set of four markets in every \acro{cat} game includes two \acro{ch} markets --- $\acro{ch}_l$ and $\acro{ch}_h$ --- and two \acro{cda} markets --- $\acro{cda}_l$ and $\acro{cda}_h$ --- with one of each charging $10\%$ on trader profit, like $\ar{gf}_{0.1}$ does, and the other charging $100\%$ on trader profit (denoted as $\ar{gf}_{1.0}$). \acro{ch} and \acro{cda} mechanisms are two common double auctions and have been used in the real world for many years, in financial marketplaces in particular due to their high allocative efficiency. Earlier experiments we ran, involving \acro{ch} and \acro{cda} markets against entries into \acro{cat} competitions, indicate that it is not trivial to win over these two standard double auctions. Markets with different charge levels are included to avoid any sampled mechanisms taking advantage otherwise. Based on the parameterized framework in Section~\ref{sec:greybox:gda}, the \acro{ch} and \acro{cda} markets can be represented as follows:
\begin{center}
\begin{tabular}{r@{ / }l@{ = }l}
  $\acro{ch}_l$ & $\acro{ch}_h$  & \ar{me} + \ar{qt} + \ar{aq} + \ar{cr} + $\ar{pu}_{k=0.5}$ + $\ar{gf}_{0.1}$ / $\ar{gf}_{1.0}$ \\
  $\acro{cda}_l$ & $\acro{cda}_h$ & \ar{me} + \ar{qt} + \ar{aq} + \ar{cc} + $\ar{pd}_{k=0.5}$ + $\ar{gf}_{0.1}$ / $\ar{gf}_{1.0}$
\end{tabular}
\end{center}

The Hall of Fame that we maintain during the search contains ten `active' members and a list of `inactive' members. After each \acro{cat} game, the two sampled mechanisms are compared with those active Hall of Famers. If the score of a sampled mechanism is higher than the lowest average score of the active Hall of Famers, the sampled mechanism is inducted into the Hall of Fame and replaces the corresponding Hall of Famer, which becomes inactive and ineligible for \acro{cat} games at later steps. An inactive Hall of Famer may be reactivated if an identical mechanism happens to be sampled from the space again and scores high enough to promote its average score to surpass the lowest score of active Hall of Famers.

Each \acro{cat} game is populated by 120 trading agents, using \acro{zi-c}, \acro{zip}, \acro{re}, and \acro{gd} strategies, a quarter of the traders using each strategy. Half the traders are buyers, half are sellers. The supply and demand schedules are both drawn from a uniform distribution between 50 and 150. Each \acro{cat} game lasts 500 days with ten rounds for each day. This setup is similar to that of actual \acro{cat} competitions except for a smaller trader population that helps to reduce computational costs. A 200-step grey-box experiment takes around sixteen hours on a \acro{windows pc} that runs at 2.8GHz and has a 3GB memory.

\subsection{Experimental results}
\label{sec:experiments:results}

We carried out four experiments to check whether the grey-box approach is successful in searching for good auction mechanisms.

\begin{figure}[tb]
\begin{center}
\subfloat[The four fixed auction mechanisms.]{\label{fig:fixed-markets}\includegraphics[scale=0.4]{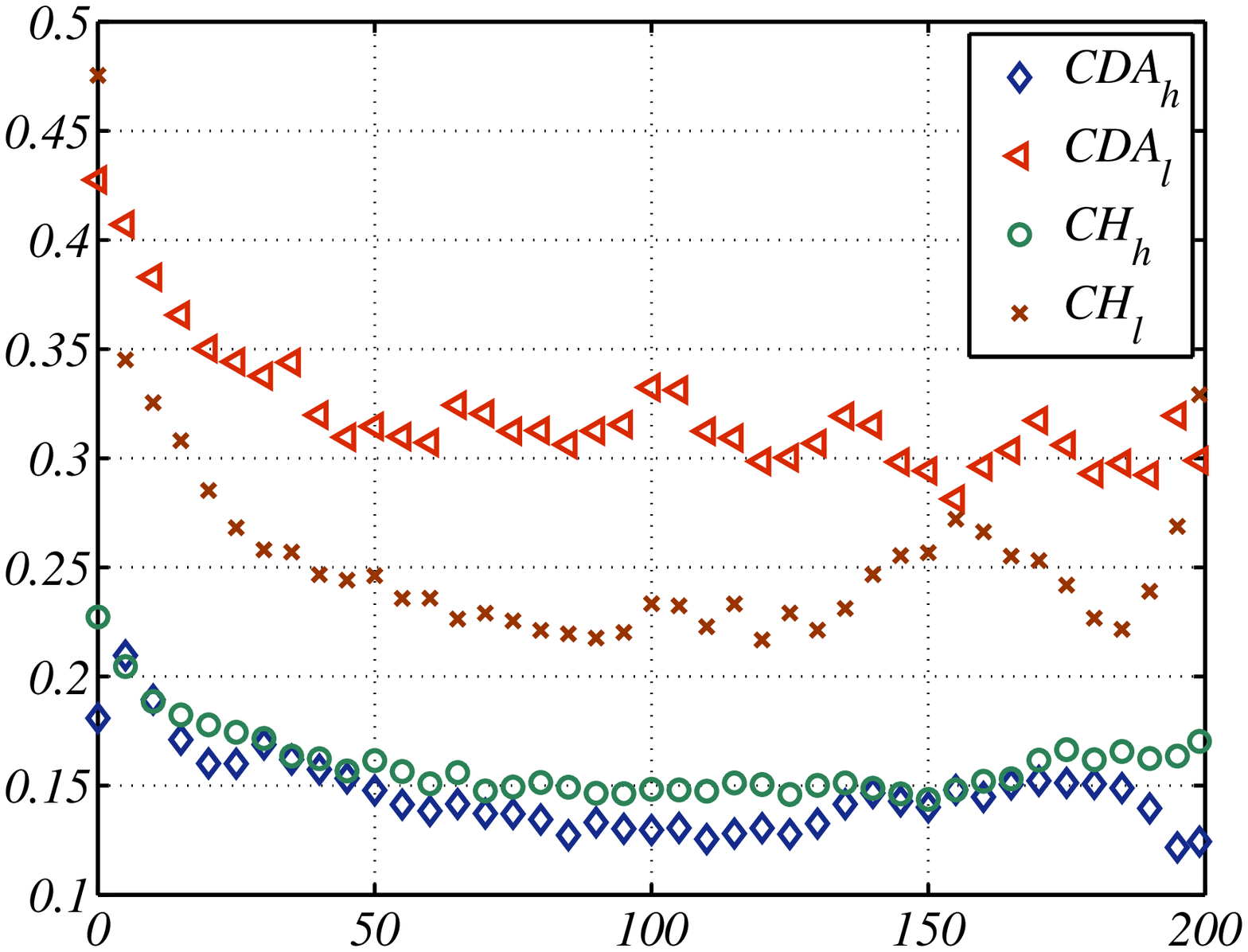}}
\qquad
\subfloat[The top ten active Hall of Famers.]{\label{fig:elite}\includegraphics[scale=0.4]{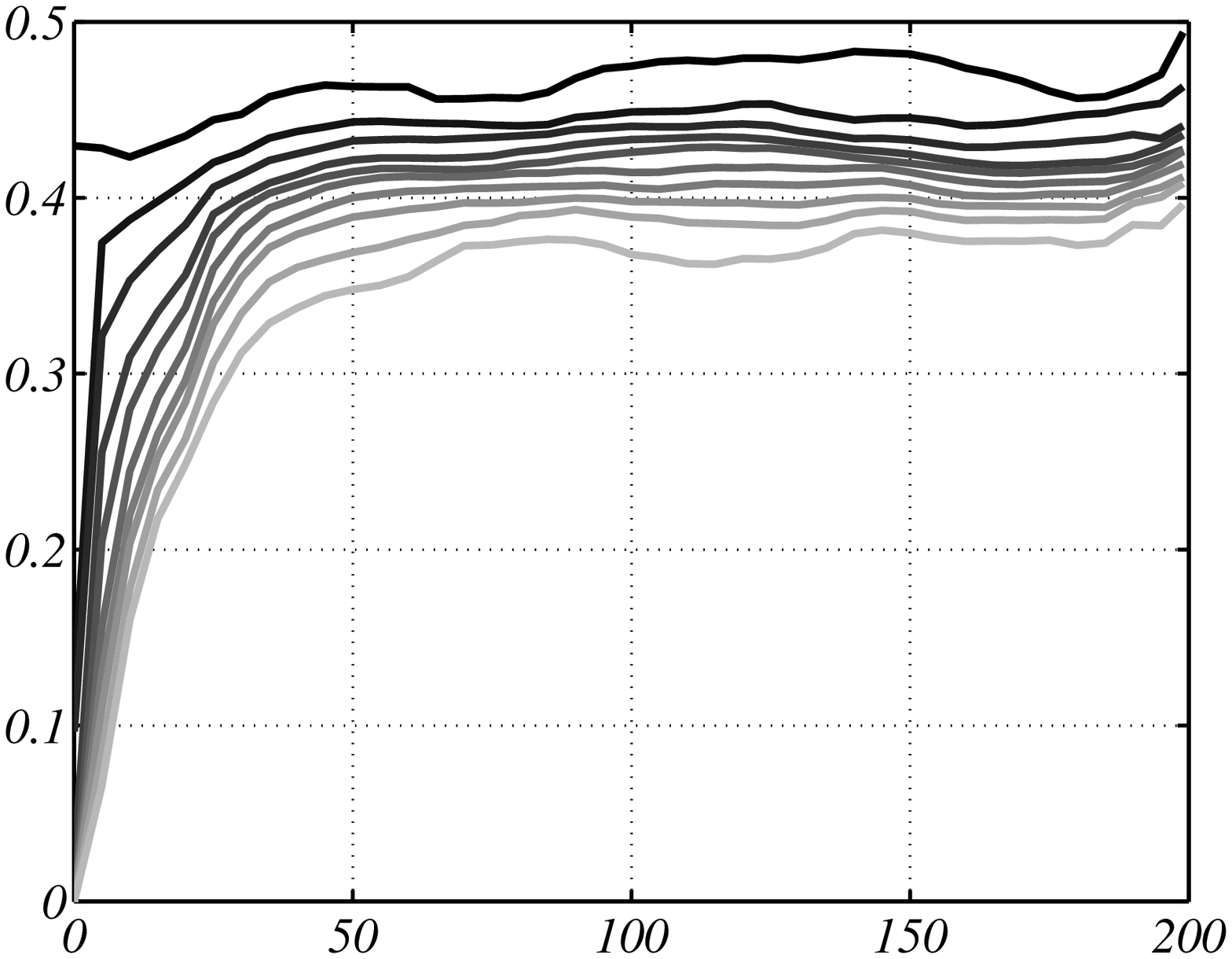}}
\caption{Scores of market mechanisms over 200 steps during the grey-box process.}
\label{fig:scores}
\end{center}
\end{figure}

First, we measured the performance of the mechanisms that are being generated indirectly, through their effect on other mechanisms. Since the four standard markets participate in all the \acro{cat} games, their performance over time reflects the strength of their opponents --- they will do worse as their opponents get better --- which in turn reflects whether the search generates increasingly better mechanisms. Figure~\ref{fig:fixed-markets} shows that the scores of the four markets (more specifically the average daily scores of the markets in a game) decrease over 200 games, especially over the first 100 games, suggesting that the mechanisms we are creating get better as the learning process progresses.

Second, we measured the performance of the set of mechanisms we are creating more directly. The mechanisms that are active in the Hall of Fame at a given point represent the best mechanisms that we know about at that point and their performance tells us more directly how the best mechanisms evolve over time. Figure~\ref{fig:elite} shows the scores of the ten active Hall of Famers at each step over a 200-step run.\fn{Note that the active Hall of Famers will be different mechanisms at different steps in the process, so what we see in the figure is the performance of the best mechanisms we know of.} As in Figure~\ref{fig:fixed-markets}, the first 100 steps sees a clear, increasing trend. Note that even the scores of the worst of the ten at the end are above 0.35, higher than the highest of the four fixed markets from Figure~\ref{fig:fixed-markets}. Thus we know that our approach will create mechanisms that outperform standard mechanisms, though we should not read too much into this since we trained our new mechanisms directly against them.

Third, a better test of the new mechanisms is to run them against those mechanisms that we know to be strong in the context of \acro{cat} games, asking what would have happened if our Hall of Fame members had been entered into prior \acro{cat} competitions and had run against the carefully hand-coded entries in those competitions. We chose three Hall of Famers, which are internally labeled as \s{SM7.1}, \s{SM88.0}, and \s{SM127.1} and can be represented in the parameterized framework in Section~\ref{sec:greybox:gda} as follows:
\begin{center}
\begin{tabular}{r@{ = }l}
 \s{SM7.1} & \ar{mv} + \ar{qo} + $\ar{ah}_{\tau=0.4}$ + $\ar{cp}_{p=0.3}$ + $\ar{pn}_{n=11}$ + $\ar{gf}_{0.1}$ \\
 \s{SM88.0} & $\ar{mt}_{\theta=0.4}$ + \ar{qt} + \ar{aa} + $\ar{cp}_{p=0.4}$ + $\ar{pu}_{k=0.7}$ + $\ar{gf}_{0.1}$ \\
 \s{SM127.1} & \ar{mv} + \ar{qs} + \ar{as} + $\ar{cp}_{p=0.4}$ + $\ar{pu}_{k=0.7}$ + $\ar{gf}_{0.1}$ \\
\end{tabular}
\end{center}
We ran these three mechanisms against the best recreation of past \acro{cat} competitions that we could achieve given the contents of the \acro{tac} agent repository,\fn{\url{http://www.sics.se/tac/showagents.php}.} where competitors are asked to upload their entries after the competition. There were enough entries in the repository at the time we ran the experiments to create reasonable facsimiles of the 2007 and 2008 competitions, but there were not enough entries from the 2009 competition for us to recreate that year's competition. The \acro{cat} games were set up in a similar way to the competitions, populated by 500 traders that are evenly split between buyers and sellers and between the four trading strategies --- \acro{zi-c}, \acro{zip}, \acro{re}, and \acro{gd} --- and the private values of sellers or buyers were drawn from a uniform distribution between 50 and 150. For each recreated competition, we ran three games, like in the actual competitions.

Table~\ref{tab:catscores} lists the average cumulative scores of all the markets across their three games along with the standard deviations of those scores. The three new mechanisms we obtained from the grey-box approach beat the actual entries in the competition by a comfortable margin in both cases. The fact that we can take mechanisms that we generate in one series of games (against the fixed opponents and other new entries) and have them perform well against a separate set of mechanisms suggests that the grey-box approach learns robust mechanisms.

In passing, we note that the rankings of the entries from the repository do not reflect those in the actual \acro{cat} competitions. This is to be expected since the entries now face much stronger opponents and different markets will, in general, respond differently to this. Excluding the markets that attempt to impose invalid fees and are marked with `*', we can see that the overall performance of entries into the 2008 \acro{cat} competition is better than that of those into the 2007 \acro{cat} competition when they face the three new, strong, opponents, reflecting the improvement in the entries over time.
%
\begin{table*}[tb]
\begin{center}
\caption{The scores of markets in \acro{cat} games including the best mechanisms from the grey-box approach and entries in prior \acro{cat} competitions, averaged over three \acro{cat} games respectively.}
\label{tab:catscores}
  \subfloat[Against \acro{cat} 2007 entries.]{\label{tab:catscores-07}
    \begin{tabular}{+lrr}
    \toprule
    \textbf{Market}  & \textbf{Score} & \textbf{\acro{sd}}\\
    \otoprule
    \s{SM7.1} & 199.4500 & 5.9715 \\
    \s{SM88.0} & 191.1083 & 10.3186 \\
    \s{SM127.1} & 180.1277 & 9.0289 \\
    \s{MANX} & 154.6953 & 1.3252 \\
    \s{CrocodileAgent} & 142.0523 & 9.0867 \\
    \s{TacTex} & 138.4527 & 5.8224 \\
    \s{PSUCAT} & 133.1347 & 5.6565 \\
    \s{PersianCat} & 124.3767 & 11.2409 \\
    \s{jackaroo} & 108.8017 & 8.6851 \\
    \s{IAMwildCAT}\tmark[*] & 106.8897 & 4.4006 \\
    \s{Mertacor} & 89.1707 & 4.9269 \\
    \ \\
    \bottomrule
    \end{tabular}
  }
  \qquad
  \qquad
  \qquad
  \subfloat[Against \acro{cat} 2008 entries.]{\label{tab:catscores-08}
    \begin{tabular}{+lrr}
    \toprule
    \textbf{Market}  & \textbf{Score} & \textbf{\acro{sd}}\\
    \otoprule
    \s{SM7.1} & 196.7240 & 9.2843 \\
    \s{SM88.0} & 186.9247 & 4.2184 \\
    \s{SM127.1} & 183.5887 & 9.7835 \\
    \s{jackaroo} & 177.5913 & 2.5722 \\
    \s{Mertacor} & 161.5440 & 5.8741 \\
    \s{MANX} & 147.3050 & 15.7718 \\
    \s{IAMwildCAT} & 142.9167 & 8.9581 \\
    \s{PersianCat} & 139.1553 & 17.9783 \\
    \s{DOG} & 130.2197 & 18.9782 \\
    \s{MyFuzzy} & 125.9630 & 1.9221 \\
    \s{CrocodileAgent}\tmark[*] & 71.4820 & 5.8687 \\
    \s{PSUCAT}\tmark[*] & 68.3143 & 6.7389 \\
    \bottomrule
    \end{tabular}
  }

\begin{minipage}{0.99\linewidth}
{\scriptsize{*}} \s{IAMwildCAT} from \acro{cat} 2007,  and \s{CrocodileAgent} and \s{PSUCAT} from \acro{cat} 2008 worked abnormally during the games and tried to impose invalid fees, probably due to competitions from the three new, strong opponents. Although we modified \acro{jcat} to avoid kicking out these markets on those trading days when they impose invalid fees --- which \acro{jcat} does in an actual \acro{cat} tournament --- these markets still perform poorly, in contrast to their rankings in the tournaments.
\end{minipage}
\end{center}
\end{table*}

Finally, we tested the performance of \s{SM7.1}, \s{SM88.0}, and \s{SM127.1} when they are run in isolation, applying the same kind of test that auction mechanisms are traditionally subject to.
We tested the mechanisms both for allocative efficiency and, following \cite{niu-cps-06-aamas-pricing.fluctuation}, for the extent to which they trade close to theoretical equilibrium as measured by the coefficient of convergence, $\alpha$. Niu \textit{et al.} \cite{niu-cps-06-aamas-pricing.fluctuation} compared a class of double auctions, called \acro{ncdaee}, which can be represented as:
\begin{center}
\acro{ncdaee} = \ar{me} + $\ar{ae}_{w,\delta}$ + \ar{cc} + $\ar{pn}_n$
\end{center}
The advantage of \acro{ncdaee} is that it can give significantly lower $\alpha$ --- faster convergence of transaction prices --- and higher allocative efficiency ($E_a$) than a \acro{cda} when populated respectively by homogeneous \acro{zi-c} traders and can perform comparably to a \acro{cda} when populated by homogeneous \acro{gd} traders.

We replicated these experiments using \acro{jcat} and ran additional ones for the three new mechanisms with similar configurations. The results of these experiments are shown in Table~\ref{tab:economic}.\fn{Our results are slightly different from those in \cite{niu-cps-06-aamas-pricing.fluctuation}, but the pattern of these results still holds. In addition, we ran an \acro{ncdaee} variant ($\delta=30$) that was not tested in \cite{niu-cps-06-aamas-pricing.fluctuation}, observing that those with $\delta\leq 20$ do not perform well when populated by \acro{gd} traders.} The best result in each column is shaded. We can see that both cases of \s{SM7.1} with \acro{zi-c} traders and \s{SM88.0} with \acro{gd} traders give higher $E_a$ than the best of the existing markets respectively, and both of these increases are statistically significant at the 95\% level. Both cases also lead to low $\alpha$, not the lowest in the column but close to the lowest, and the differences between them and the lowest are not statistically significant at the 95\% level. Thus the grey-box approach can generate mechanisms that perform as well in the single market scenario as the best mechanisms from the literature.
\begin{table*}[tb]
\begin{center}
\caption{Economic properties of the best mechanisms from the grey-box experiments and the auction mechanisms explored in \protect\cite{niu-cps-06-aamas-pricing.fluctuation}. All \acro{ncdaee} mechanisms are configured to have $w=4$ in their \ar{ae} policies and $n=4$ in their \ar{pn} policies.The best result in each column is shaded. Data in the first four row are averaged over 1,000 runs and those in the last four are averaged over 100 runs.}
\label{tab:economic}
    \begin{tabular}{+lrrrrrrrr}
    \toprule\rowstyle{\centering}
    \multirow{3}*{\textbf{Market}} & \multicolumn{4}{c}{\textbf{\acro{zi-c}}} & \multicolumn{4}{c}{\textbf{\acro{gd}}} \\ \cmidrule (l){2-5} \cmidrule (l){6-9}
    & \multicolumn{2}{c}{\textbf{$E_a$}} & \multicolumn{2}{c}{\textbf{$\alpha$}} & \multicolumn{2}{c}{\textbf{$E_a$}} & \multicolumn{2}{c}{\textbf{$\alpha$}} \\ \cmidrule (l){2-3} \cmidrule (l){4-5} \cmidrule (l){6-7} \cmidrule (l){8-9}
    & \textbf{Mean} & \textbf{\acro{sd}} & \textbf{Mean} & \textbf{\acro{sd}} & \textbf{Mean} & \textbf{\acro{sd}} & \textbf{Mean} & \textbf{\acro{sd}} \\
    \otoprule
    \acro{cda} & 97.464 & 3.510 & 13.376 & 4.351 & 99.740 & 1.553 & \cellcolor[gray]{0.9}4.360 & 3.589 \\
    $\acro{ncdaee}_{\delta=0}$ & 98.336 & 3.262 & \cellcolor[gray]{0.9}4.219 & 3.141 & 9.756 & 28.873 & 14.098 & 1.800 \\
    $\acro{ncdaee}_{\delta=10}$ & 98.912 & 2.605 & 5.552 & 2.770 & 23.344 & 41.727 & 7.834 & 5.648 \\
    $\acro{ncdaee}_{\delta=20}$ & 98.304 & 2.562 & 7.460 & 3.136 & 89.128 & 30.867 & 4.826 & 3.487 \\
    $\acro{ncdaee}_{\delta=30}$ & 97.708 & 3.136 & 8.660 & 3.740 & 99.736 & 1.723 & 4.498 & 3.502 \\ \midrule
    \s{SM7.1} & \cellcolor[gray]{0.9}99.280 & 1.537 & 4.325 & 2.509 & 58.480 & 47.983 & 4.655 & 4.383 \\
    \s{SM88.0} & 98.320 & 2.477 & 11.007 & 4.251 & \cellcolor[gray]{0.9}99.920 & 0.560 & 4.387 & 2.913 \\
    \s{SM127.1} & 97.960 & 3.225 & 11.152 & 4.584 & 99.520 & 1.727 & 4.751 & 3.153 \\
    \bottomrule
    \end{tabular}
\end{center}
\end{table*}

\section{Conclusions and future work}
\label{sec:conclusions}

This paper describes a practical approach to the automated design of complex mechanisms. The approach that we propose breaks a mechanism down into a set of components each of which can be implemented in a number of different ways, some of which are also parameterized. Given a method to evaluate candidate mechanisms, the approach then uses machine learning to explore the space of possible mechanisms, each composed from a specific choice of components and parameters. The key difference between our approach and previous approaches to this task is that the score from the evaluation is not only used to grade the candidate mechanisms, but also the components and parameters, and new mechanisms are generated in a way that is biased towards components and parameters with high scores.

The specific case-study that we used to develop our approach is the design of new double auction mechanisms. Evaluating the candidate mechanisms using the infrastructure of the TAC Market Design competition, we showed that we could learn mechanisms that can outperform the standard mechanisms against which learning took place and the best entries in past Market Design competitions. We also showed that the best mechanisms we learnt could outperform mechanisms from the literature even when the evaluation did not take place in the context of the Market Design game. These results make us confident that we can generate robust double auction mechanisms and, as a consequence, that the grey-box approach is an effective approach to automated mechanism design.

Now that we can learn mechanisms effectively, we plan to adapt the approach to also learn trading strategies, allowing us to co-evolve mechanisms and the traders that operate within them.

\section*{Acknowledgments}

This work is partially supported by the \acro{US NSF} under grant IIS-0329037, \emph{Tools and
Techniques for Automated Mechanism Design}, and the \acro{UK EPSRC} under grants GR/T10657/01 and GR/T10671/0, \emph{Market Based Control of Complex Computational Systems}. We thank the entrants to the Market Design game for releasing binaries of their market agents.



\end{document}